# Spin Jam: a quantum-fluctuation-induced glassy state of a frustrated magnet

J. Yang[a], A. M. Samarakoon[a], S. E. Dissanayake[a], H. Ueda[b], I. Klich[a], K. Iida[c], D. Pajerowski[d], N. Butch[d], Q. Huang[d], J.R.D. Copley[d], S.-H. Lee[a,1]

[a] Department of Physics, University of Virginia, Charlottesville, Virginia 22904, USA

[b] Division of Chemistry, Graduate School of Science, Kyoto University, Kyoto 606-8502, Japan

[c] Research Center for Neutron Science and Technology, Comprehensive Research Organization for Science and Society (CROSS), Tokai, Ibaraki 319-1106, Japan

[d] National Institute of Standards and Technology, Gaithersburg, Maryland 20899, USA

[1] To whom correspondence should be addressed. E-mail: shlee@virginia.edu

**Abstract:** Since the discovery of spin glasses in dilute magnetic systems, their study has been largely focused on understanding randomness and defects as the driving mechanism. The same paradigm has also been applied to explain glassy states found in dense frustrated systems. Recently, however, it has been theoretically suggested that different mechanisms, such as quantum fluctuations and topological features, may induce glassy states in defect-free spin systems, far from the conventional dilute limit. Here we report experimental evidence for existence of a glassy state, that we call a spin jam, in the vicinity of the clean limit of a frustrated magnet, which is insensitive to a low concentration of defects. We have studied the effect of impurities on $SrCr_{9p}Ga_{12-9p}O_{19}$ (SCGO($p$)), a highly frustrated magnet, in which the magnetic $Cr^{3+}$ (s=3/2) ions form a quasi-two-dimensional triangular system of bi-pyramids. Our experimental data shows that as the nonmagnetic $Ga^{3+}$ impurity concentration is changed, there are two distinct phases of glassiness: a distinct exotic glassy state, which we call a "spin jam", for high magnetic concentration region ($p > 0.8$) and a cluster spin glass for lower magnetic concentration, ($p < 0.8$). This observation indicates that a spin jam is a unique vantage point from which the class of glassy states in dense frustrated magnets can be understood.

*Significance:* *We report experimental evidence for a glassy state induced by quantum fluctuations, a spin jam, that is realized in $SrCr_{9p}Ga_{12-9p}O_{19}$ (SCGO(p)), a highly frustrated magnet, in which the magnetic $Cr^{3+}$ (s=3/2) ions form a quasi-two-dimensional triangular system of bi-pyramids. Our new experimental data and our theoretical spin jam model provide for the first time a coherent understanding of the existing experimental data of this fascinating system. Furthermore, our findings strongly support the possible existence of purely topological glassy states.*

Understanding glassy states found in dense frustrated magnets has been an intellectual challenge since peculiar low temperature glassy behaviors were observed experimentally in the quasi-two-dimensional SCGO (1-3) and in the three-dimensional (3D) pyrochlore $Y_2Mo_2O_7$ (4). Immediately following, theoretical investigations (5-9) were performed to see if an intrinsic spin freezing transition is possible in a defect-free situation, aided by quantum fluctuations, as in the order-by-fluctuations phenomenon (10,11). Quantum fluctuations at $T = 0$ were shown to select a long range ordered state in the two-dimensional (2D) kagome isotropic antiferromagnet (AFM) (5-6), later expanded to the isotropic pyroclore and SCGO (9). Anisotropic interactions were also considered as a possible origin of the glassy kagome AFM (7). For an XY pyrochlore AFM, thermal fluctuations were found to induce a conventional Neel order (8). Experimental works were also performed to investigate if the glassy states are extrinsic due to site defects, random couplings or intrinsic to the magnetic lattice (12,13). The consensus is that the low temperature spin freezing transitions in SCGO($p$) near the clean limit ($p \sim 1$) is not driven by site defects. (13)

The nature of the frozen state in SCGO has been investigated by numerous experimental techniques, including bulk susceptibility (1-3), specific heat (2,14), $\mu$SR (15), NMR (13,16), and elastic and inelastic neutron scattering (17). Observed are spin glassy behaviors, such as field-cooled and zero-field-cooled (FC/ZFC) hysteresis in bulk susceptibility (3), as well as non-spin-glassy behaviors, such as a quadratic behavior of specific heat at low $T$, $C_v \propto T^2$ (14), linear dependence of the imaginary part of the dynamic susceptibility at low energies, $\chi''(\omega) \propto \omega$ (17), and a broad but prominent momentum dependence of the elastic neutron scattering intensity (17). The interpretation of the frozen state below $T_f$ is still controversial. One possibility suggested was a spin liquid with unconfined spinons or resonating valence bond state, based on NMR and $\mu$SR studies (15,16). Many-body singlet excitations were also suggested to be responsible for the $C_v \propto T^2$ behavior (14).

Recently, some of us presented an alternative scenario involving a spin jam state by considering the effects of quantum fluctuations in the disorder-free quasi-two-dimensional ideal SCGO lattice with a simple nearest neighbor spin interaction Hamiltonian $\mathcal{H} = J \sum_{NN} S_i \cdot S_j$ (18,19). The spin jam framework provided a qualitatively coherent understanding of all the low temperature behaviors such as a complex energy landscape is responsible for the frozen state without long range order (18), and Halperin-Saslow like modes for the $C_v \propto T^2$ and $\chi''(\omega) \propto \omega$ behaviors (5,18). In this system, which we refer to as the ideal SCGO model (iSCGO), semi-classical magnetic moments (or spins) are arranged in a triangular network of bi-pyramids and interact uniformly with their nearest neighbors (18,19). The microscopic mechanism for the spin jam state is purely quantum mechanical. The system has a continuous and flat manifold of ground states at the mean field level, including locally collinear, coplanar and non-coplanar spin arrangements. Quantum fluctuations lift the classical ground state degeneracy ("order-by-fluctuations"), resulting in a complex rugged energy landscape that has a plethora of local minima consisting of the locally collinear states separated from each other by potential barriers (18). Although the work of Ref. (18) dealt with a similar phase space constriction by quantum fluctuations as the aforementioned other theoretical works did, we would like to stress here the difference between the two: While the other works mainly focused on the selection of the long-range-ordered (LRO) energetic ground state, the work of Ref. (18) showed that the short-range-ordered (SRO) states that exist at higher energies are long lived, dominate entropically over the LRO states and govern the low-$T$ physics.

The introduction of non-magnetic impurities into a topological spin jam state breaks some of the constraints in the system, and possibly allows local transitions between minima, with a time scale dependent on the density of impurities. At a sufficiently high vacancy concentration, the system exits the spin jam state and becomes either paramagnetic or an ordinary spin glass at lower temperatures. Here we try to identify and explore the spin jam regime in an experimentally accessible system. The three most important signatures we seek for the existence of a spin jam state, different from conventional spin glass states, are (a) a linear dependence of the imaginary part of the dynamic susceptibility at low energies, $\chi''(\omega) \propto \omega$, (b) intrinsic short range static spin correlations, and (c) insensitivity of its physics to nonmagnetic doping near the clean limit. In the rest of the paper we provide experimental demonstration of these properties.

Experimentally, there are so far two materials, $SrCr_{9p}Ga_{12-9p}O_{19}$ (SCGO($p$)) (1-3,13-17,20) and qs-ferrites like $Ba_2Sn_2ZnGa_3Cr_7O_{22}$ (BSZGCO) (21), in which the magnetic $Cr^{3+}$ ($3d^3$) ion surrounded by six oxygen octahedrally, form distorted quasi-two-dimensional triangular lattice of bi-pyramids (20,21) as shown in Fig. 1 (a), and thus may realize a spin jam state. We would like to emphasize that these systems are very good insulators (resistivity $\rho > 10^{13}$ $\Omega \cdot cm$ at 300 K) and the $Cr^{3+}$ $\left(t_{2g}^3\right)$ ion has no orbital degree of freedom. Furthermore, the neighboring Cr ions share one *edge* of oxygen octahedra and thus the direct overlap of the $t_{2g}^3$ orbitals of the neighboring $Cr^{3+}$ ions make the antiferromagnetic nearest neighbor (NN) Heisenberg exchange interactions dominant and further neighbor interactions negligible (22,23), as found in $Cr_2O_3$ (24) and $ZnCr_2O_4$ (25).

**Experimental data**

We have performed elastic and inelastic neutron scattering that directly probe spin-spin correlations and bulk susceptibility measurements on SCGO($p$) with various values of $p$ over $0.2 \lesssim p \lesssim 1.0$ spanning almost the entire region of $p$. To first characterize the samples and to construct the *T-p* phase diagram, we have performed dc magnetic susceptibility, $\chi_{bulk}$, and elastic neutron measurements. The data obtained from two samples with $p = 0.968(6)$ and $0.620(8)$ both of which are well above the percolation threshold for the triangular lattice of bi-pyramids, $p_c \approx 0.5$, (9) are shown in Fig. 2 (a) and (b), respectively (the data obtained from samples with other $p$ values are shown in the Supporting Information (*SI*) Fig. S3). For both samples, $\chi_{bulk}$ exhibits similar field-cooled and zero-field-cooled hysteresis below $T_f = 3.68$ K ($p = 0.968(6)$) and 1.76 K ($p = 0.620(8)$) that are much lower than their large Curie-Weiss temperatures $|\Theta_{CW}| = 504$ K ($p = 0.968(6)$) and 272 K ($p = 0.620(8)$) (see Fig. S1). The high frustration index $f = \frac{|\Theta_{CW}|}{T_f} \gtrsim 130$ indicates the presence of strong frustration in both systems. Elastic neutron scattering intensity with an instrumental energy resolution of $|\hbar\omega| \leq 25$ μeV starts developing at temperatures higher than their $T_f$ determined by $\chi_{bulk}$, as is expected for spin freezing for measurements with different energy resolution (26). Fig. 1 (b) summarizes the results obtained from all the samples studied.

We observe strikingly different behaviors in the low energy spin dynamics of the $p = 0.968(6)$ and $0.620(8)$ despite the similar temperature dependences of $\chi_{bulk}$ and of elastic neutron scattering intensity. As shown in Fig. 2 (c), in the frozen state of $p = 0.968(6)$, the inelastic neutron scattering intensity $I(Q, \hbar\omega)$ is weak at very low energies below 0.25 meV and gets stronger as $\hbar\omega$ increases. In contrast, in the frozen state of $p = 0.620(8)$, $I(Q, \hbar\omega)$ is stronger at very low energies below 0.25 meV and gets weaker as $\hbar\omega$ increases (see Fig. 2 (d)). This stark difference hints that the frozen states of $p = 0.968(6)$ and $0.620(8)$ are different in nature.

To investigate more carefully the difference between the low and high doping regions, we have also performed time-of-flight neutron scattering measurements on several samples with various doping concentration from $p = 0.968(6)$ to $0.459(5)$ (see Fig. S4). The data exhibits a continuum spectrum centered in the vicinity of $Q_{max} \approx 1.5$ Å$^{-1}$ that corresponds to (2/3,2/3,1.8), confirming that the kagome-triangular-kagome trilayer is responsible for the low energy dynamic spin correlations.(19) This resembles the energy continuum expected for spin liquids or cooperative paramagnets; however, in contrast to a spin liquid, in our system static spin correlations develop below $T_f$ as well. The $Q$-dependence of the elastic magnetic neutron scattering intensity, $I_{el}^{mag}(Q) = I_{el}(Q, 1.4K) - I_{el}(Q, 20K)$, was obtained with an energy window of $|\hbar\omega| \leq 25$ µeV, where the subtraction of the signal above the magnetic transition eliminates background from sources not related to the transition. As shown in Fig. 3, for $p = 0.968(6)$, very close to the clean limit, $I_{el}^{mag}(Q)$ exhibits a broad peak at $Q_{max} = 1.49$ Å$^{-1}$. Its broadness indicates the short-range nature of the static spin correlations. Surprisingly, upon increasing nonmagnetic impurity concentration up to about 20% ($p \sim 0.8$), the shape of $I_{el}^{mag}(Q)$ remains the same. Only upon further doping, $I_{el}^{mag}(Q)$ becomes broader, and the peak position $Q_{max}(p)$ starts shifting down.

To quantitatively investigate the correlation length $\xi(p)$ and peak position $Q_{max}(p)$ as a function of $p$, we fit $I_{el}^{mag}(Q)$ to a simple Lorentzian function for each $p$, $I_{el}^{mag}(Q) \propto \frac{1}{HWHM^2+(Q-Q_{max})^2}$ where HWHM is the half-width-half-maximum of $I_{el}^{mag}(Q)$. As a measure of the static spin correlation length scale of the frozen state, we can use $\xi_{HWHM}^{powder} = \frac{1}{HWHM}$. Note that the powder averaging would introduce extrinsic broadening to $I_{el}^{mag}(Q)$, and thus $\xi_{HWHM}^{powder}$ is underestimated, in comparison to the correlation length determined from single crystal data, $\xi_{HWHM}^{crystal} = 4.6(2)$ Å for SCGO($p \sim 0.67$) (19). But the $p$-dependence of $\xi_{HWHM}^{powder}$ serves our search for an intrinsic state.

The resulting $\xi_{HWHM}^{powder}$ and $Q_{max}$ are shown in Fig. 3b. Remarkably, both

$\xi_{HWHM}^{powder}$ and $Q_{max}$ exhibit a flat behavior near the clean limit up to 20 % doping, which is a direct evidence for existence of a distinct phase over $1 < p < 0.8$ where the *intrinsic* short range static correlations are independent of nonmagnetic doping. This can be naturally described as a spin jam state at the clean limit. A spin jam state can be intrinsically short range, and has an intrinsic static correlation length, $\xi_{int}$. Thus, when nonmagnetic doping is low and the typical distance between the nonmagnetic impurities is larger than $\xi_{int}$ the spin correlations do not get affected. Only upon significant doping, the spin correlations would get disturbed to make $\xi(p)$ shorter than $\xi_{int}$ when typical distance between impurities is short.

If the $p > 0.8$ phase is distinct from the lower $p < 0.8$ phase, then the low energy spin excitations in those two phases should have different characteristics. To see this, we show in Fig. 4 the imaginary part of the dynamic susceptibility $\chi''(\hbar\omega) = \int_{0.4 \text{ Å}^{-1}}^{1.8 \text{ Å}^{-1}} \chi''(Q, \hbar\omega) dQ$ where $\chi''(Q, \hbar\omega) = \pi I(Q, \hbar\omega)\{1 - e^{-\hbar\omega/k_B T}\}$ and $k_B$ is the Boltzmann factor. For $p = 0.968(6)$, upon cooling in the cooperative paramagnetic state from 20 K ($\ll |\Theta_{CW}|$) to 5 K ($\sim T_f$) $\chi''(\hbar\omega)$ changes from being linear to almost flat (see Fig. 4 (a)). The nearly $\omega$-independent low energy $\chi''$ implies that a distribution of the characteristic spin relaxation rates, $\Gamma$, is present, which is common in glassy transitions. (27) A distribution of $\Gamma$ would yield $\chi''(\omega) \propto \int_{\Gamma_{min}}^{\Gamma_{max}} \frac{\omega}{\omega^2 + \Gamma^2} d\Gamma = \tan^{-1}\frac{\omega}{\Gamma_{min}}$ where a term, $\tan^{-1}\frac{\omega}{\Gamma_{max}}$, is ignored since $\Gamma_{max} \gg 1$ meV, much larger than the $\omega$-range of interest. Fitting the $T = 6$ K $\sim T_f$ data to the model yields a distribution of $\Gamma$ with $\Gamma_{min} = 0.053(4)$ meV (see the red solid line in Fig. 4 (a)).

Upon further cooling into the frozen state, however, $\chi''(\hbar\omega)$ exhibits hardening: the weight gets depleted and becomes linear at low energies (see the 1.4 K data), which is consistent with a previous neutron scattering study of SCGO($p = 0.92(5)$) (17). The 1.4 K data can still be fitted to $\chi''(\omega) \propto \tan^{-1}\frac{\omega}{\Gamma_{min}}$ with $\Gamma_{min} = 0.25(3)$ meV, which indicates that a distribution of $\Gamma$, is still present, but with the larger minimum cutoff than that of 6 K. And for $\hbar\omega \lesssim 2\Gamma_{min}$, $\chi''(\omega)$ is linear with $\omega$.

For $p = 0.459(5)$, on the other hand, we observe a fundamentally different behavior in the frozen state: as shown in Fig. 4(f), below freezing, rather than hardening, the spectral weight at low frequencies $\chi''(\hbar\omega)$ exhibits a prominent increase peaked at ~ 0.2 meV. Obviously, the data cannot be reproduced by $\tan^{-1}\frac{\omega}{\Gamma_{min}}$ alone. Instead it behaves more like a Lorentzian with one characteristic relaxation rate, $\chi''(\omega) \propto \frac{\omega}{\omega^2+\Gamma_L^2}$.

The clear difference in the behavior of $\chi''(\hbar\omega)$ in the frozen state between the two regimes of $p$ is another evidence that the frozen state in the vicinity of the clean limit is indeed a distinct state. In order to see how the evolution of the state occurs as a function of $p$, we have fitted $\chi''(\omega)$ measured in the frozen state of each $p$ to the sum of the two contributions, $\chi''(\omega) \propto \left(\tan^{-1}\frac{\omega}{\Gamma_{min}} + f \cdot \frac{\omega}{\omega^2+\Gamma_L^2}\right)$. The black solid lines in Fig. 4 (a)-(f) are the fitting results of the sum, while the dotted and dashed lines are the contributions of arctan and Lorentzian, respectively. The fraction of the contribution from the single Lorentzian to $\chi''(\omega)$, $f$, was determined in the vicinity to the pure limit to be zero; $f = 0.00(6)$ for $p = 0.968$ and $f = 0.00(5)$ for $p = 0.917$. As $p$ decreases further, i.e., nonmagnetic doping increases, however, $f$ gradually increases. As shown in Fig. 4 (c) and (d), the $\tan^{-1}$ term (dotted line) dominates for $p \gtrsim 0.777(6)$, while the Lorentzian term (dashed line) dominates for $p \lesssim 0.7$. $f(p)$ over a wide range of $p$ is plotted in Fig. 3 (c). This confirms, upon doping, a crossover from a frozen state near the clean limit to another frozen state at high nonmagnetic concentration limit.

What is then the nature of the frozen state in the vicinity of the clean limit? The hardening and linear behavior of the low energy spin fluctuations in the frozen state of SCGO($p > 0.8$) can be explained as Halperin-Saslow (HS) type modes in a spin jam (5,28,29). HS modes are long-range collective modes which may be viewed as an analogue of the Goldstone modes associated with continuous symmetry breaking in systems without long range order (5). In contrast, for low values of $p$, as defect concentration is increased, the low energy spectral weight is eventually dominated by contributions from a distribution of local spin clusters. This is consistent with the previous specific heat data that reported the $C_v \propto T^2$ behavior robust against dilution for $p \gtrsim 0.8$ (14), while for $p \lesssim 0.8$ the exponent starts decreasing with decreasing $p$ (30).

As a further quantitative evidence of a HS mechanism for the specific heat, $C_v/T^2$ may be roughly estimated for a spin jam. HS theory with a dispersion $\omega_k \cong \gamma \sqrt{\frac{\rho_s}{\chi_0}} k$, gives a contribution to specific heat $\frac{C_v}{(k_B T)^2} \cong \frac{9\zeta(3) k_B \chi_0}{\pi \hbar^2 \rho_s \gamma^2}$, where $\gamma$ is the gyromagnetic ratio, $\zeta(3) \sim 1.2$ is the Riemann Zeta function, and we have accounted for 3 modes for isotropic spins. Using experimental values of $\chi_0 \cong 0.007$ emu/moleCr (see Fig. 2 (a)) and using the HS upper bound estimate $\rho_s \cong -\frac{J}{18 V} \sum_{n.n.} r_{ij}^2 \langle S_i \cdot S_j \rangle$ adapted to iSCGO, $J \sim 9$meV and $\rho_s \cong -\frac{2J d^2}{18 V_{unit\ cell}} z \langle S^2 \rangle$, taking $|\langle S \rangle| = 0.95 \mu_B/Cr$ from Ref. (17), we get a theoretical estimate of $\frac{C_v}{T^2} \simeq 0.07 \frac{Joule}{moleCr\ K^3}$, which is consistent with the experimental value of $\sim 0.059 \frac{Joule}{moleCr\ K^3}$ (14).

## Discussion

The salient features observed in SCGO($p > 0.8$) are consistent with the recent understanding of the iSCGO case with $J = J'$ spin jam (18), where $J$ and $J'$ are the intra and inter-layer Heisenberg nearest neighbor couplings, respectively. For real SCGO samples $J' < J$, in general, due to lattice distortion. Furthermore, there are two different $J$s in the kagome plane; $J_1$ and $J_2$ for the bonds within each bi-pyramid and for the bonds of the linking triangle connecting the bi-pyramids. However, at the mean-field level, the case of $J_1 \neq J_2$ yields the same ground states as the uniform case of $J_1 = J_2 = J$ because the ground state of the uniform case satisfies both the AFM constraints of the linking triangle and the bi-pyramid. Thus, we limit our discussion to the case of $J_1 = J_2 = J > J'$. For these situations classical magnetic ground states have been described as a function of $J'/J$ in Ref. (19). These can be obtained from the $J = J'$ case by coherent rotations of a subset of the spins. The classical manifold of ground states remains degenerate for $J' < J$, as can be seen from the flat zero energy bands which are present in the linear spin wave analysis around long range ordered states described in Fig. S5. As the ratio $J'/J$ changes from 1 to 0, the system moves from the ideal tri-layer to decoupled kagome layers (accompanied by a layer of non interacting spins), at which point local zero energy modes, such as weathervane modes (7), associated with kagome physics, appear. For the isolated semi-classical kagome, an extensive configurational entropy of local minima appears with kinetic barriers associated with the weathervane motion. At low enough temperatures tunneling is suppressed and the system freezes. On general grounds, local modes such as the weathervane will become delocalized when the layers are coupled, suggesting that for $J \neq J' \neq 0$, as for $J = J'$, the barriers between local minima remain non-local in nature and thus freezing is more effective than for the decoupled kagome.

The quantum fluctuations-induced spin jam scenario is consistent with the system freezing at temperatures much lower than $|\Theta_{CW}|$. An energy scale for spin fluctuations in the clean limit can be determined from the potential barrier between local (ordered) minima and is given by $E_{SCGO} \approx 0.05\,JS$ (18). For SCGO( $p > 0.8$ ), $E_{SCGO}/k_B \approx 7.8$ K, which is close to the experimentally determined $T_f \approx 4$ K $\ll |\Theta_{CW}| \sim 500$ K.

One may consider anisotropic interactions such as the Dzyaloshinskii-Moriya (DM) as the origin for the spin freezing. We believe that a simple DM interaction will not generate such a complex energy landscape and cannot explain coherently all the experimental data as the quantum fluctuation-induced spin jam scenario does. However, theoretical confirmation about such an anisotropic scenario is needed, which is beyond the scope of this paper.

Upon weak non-magnetic doping the complex energy landscape is modified, the kinetic barriers become finite, however, the overall picture remains the same. Further doping will weaken the order by fluctuations mechanism and the selection of coplanar states, and a different glassy state emerges. Indeed, our observation of a crossover as a function of doping are consistent with Henley (9) who, remarkably, speculated the possibility of a defect induced crossover as a function of doping, close to the clean limit, from a non-generic phase (dominated by coplanar states) to a generic phase in large spin kagome and SCGO systems.

It is worthwhile to discuss here the concept of jamming in a broader context. In granular systems, a transition occurs into a special type of glassy state called jammed state with the increase of the density of constituents due to a complex energy landscape (31). In analogy, one may view the approach of the disorder-free frustrated magnets as the addition of system constituents, i.e. removal of non-magnetic impurities, and the state of the clean system as the "spin jammed" state. In such a system, a large set of mostly irregular spin configurations form local minima, the transition between which requires the simultaneous reorientation of a large number of spins. The recent theoretical work of Ref. (18) has shown how a disorder-free quasi-two-dimensional frustrated magnet with a simple nearest neighbor spin interaction Hamiltonian $\mathcal{H} = J\sum_{NN} S_i \cdot S_j$ can exhibit a spin jam of topological nature, at low temperatures. It was also established that the locally collinear states differ by collective reorientations of spins, where, the smallest units of the mean-field zero-energy excitations are extended along lines, called spaghetti modes. The extended nature of the reorientations is responsible for particularly large dynamical barriers, of topological nature, to tunneling between minima.(18) The effect of an "order-by-fluctuations" may also be a possible mechanism for forming a glassy state in kagome systems (5,6,32,33), with less robust nature than in the iSCGO (9), as the presence of local tunneling through weathervane modes involving 6 spins may hasten approach to a global ground state.

In comparison, surprisingly the three-dimensional pyrochlore $Y_2Mo_2O_7$ also exhibits at low temperatures a similar $C_v \propto T^D$ behavior but with $D$ less than the value of 3 expected for a three-dimensional system.(34) The magnetic $Mo^{4+}$ ($4d^2$) ions form a three-dimensional network of corner-sharing tetrahedra. If these magnetic moments are isotropic, and antiferromagnetically and uniformly interacting with their nearest neighbors only, the system is supposed to yield the highest degree of frustration. One may speculate that the freezing in $Y_2Mo_2O_7$ may also be explained in terms of a spin jam, however we would like to point out crucial differences between the systems. In particular, its frustration index is two orders of magnitude smaller than in SCGO; $f = \frac{|\Theta_{CW}|}{T_f} \approx \frac{45\,K}{20\,K} \approx 2.3.$ This can be understood by the facts that, unlike SCGO which is an excellent insulator, $Y_2Mo_2O_7$ is semi-conducting ($\rho \sim 10^{-2}\,\Omega\cdot$cm at 300 K), and the neighboring Mo ions share one *corner* of oxygen octahedra, which tends to result in non-negligble longer range magnetic interactions (22). More importantly, the magnetic Mo$^{4+}$ ($t_{2g}^2$) ions are orbitally degenerate (34,35); it is well known that orbital degeneracy has a great tendency to modify the nature of a magnetic network, and as a result it reduces dimensionality of the magnetic interactions and frustration as well, as found in ZnV$_2$O$_4$ (36). Therefore, the spin glassy state of Y$_2$Mo$_2$O$_7$ may be due to spatially random coupling constants induced by the orbital degrees of freedom rather than strong frustration. The effectively reduced dimensionality may be related to the observed $C_v \propto T^D$ with $D < 3$. A previous muon spin relaxation study (37) showed that the spin glassy state of Y$_2$Mo$_2$O$_7$ seems to remain intact at 20% of nonmagnetic doping as in SCGO, which may indicate that the jamming physics is still relevant to Y$_2$Mo$_2$O$_7$ as well. We believe that the concept of jamming and complex energy landscape can unify the seemingly different glassy states found in dense magnetic systems, such as SCGO, Y$_2$Mo$_2$O$_7$ (30), spin ice (38), and even the dynamics of magnetic domain boundaries (39).

**Conclusion**

The search for glassiness that arises intrinsically without defects and randomness has been revived recently as such glassiness may bear intricate relations with topological order (40), lack of thermalization in many body localization (41), jamming in structural glasses (42) and glassiness in super-cooled liquids (43). Our experiments indicate that quantum fluctuations, via an "order by fluctuations" mechanism, induce a glassy state, a spin jam, in the strongly frustrated SCGO($p > 0.8$), which is robust, and extends to the clean limit. The findings strongly support the possible existence of purely topological glassy states.

**Acknowledgments.** The works at University of Virginia (SHL and IK) were supported in part by the US National Science Foundation (NSF) grant DMR-1404994 and the NSF CAREER grant DMR-0956053, respectively. This work utilized facilities supported in part by the NSF under Agreement No. DMR-0944772, and was supported in part by NSF Grant No. PHYS-1066293 and the hospitality of the Aspen Center for Physics. JY and SHL thank Drs. Matthias Thede and Andrey Zheludev for their help during some of our bulk susceptibility measurements performed at ETH Zurich.

**Figure captions**

**Fig. 1** (a) In $SrCr_{9p}Ga_{12-9p}O_{19}$ (SCGO($p$)), the magnetic $Cr^{3+}$ ($3d^3$, s=3/2) ions form the kagome-triangular-kagome tri-layer (top panel). The blue and red spheres represent kagome and triangular sites, respectively. When viewed from the top of the layers, they form the triangular network of bi-pyramids (bottom panel). (b) The $p$-$T$ phase diagram of SCGO($p$) constructed by bulk susceptibility and elastic neutron scattering measurements on powder samples with various $p$ values. The freezing temperatures, $T_f$, marked with blue square and black circle symbols are obtained by bulk susceptibility and elastic neutron scattering measurements, respectively. Filled blue squares represent the data obtained from samples whose crystal structural parameters including the Cr/Ga concentrations were refined by neutron diffraction measurements (see Fig. S2 and Table S1-S4 in the Supporting Online Material), and open blue squares represent samples with nominal $p$ values. For nominal $p$ = 0.2, no freezing was observed down to 50 mK (see Fig. S4).

**Fig. 2** (a), (b) $T$-dependences of bulk susceptibility (blue solid squares) and elastic neutron scattering intensity (black open circles) for SCGO($p$) (a) with $p$ = 0.968(6) and (b) $p$ = 0.620(8). Bulk susceptibility was measured with an application of external magnetic field $H$ = 0.01 Tesla. The black dash lines indicate the nonmagnetic background for the elastic neutron scattering. (c), (d) Contour maps of neutron scattering intensity as a function of momentum and energy transfer  for (c) $p$ = 0.968(6) and (d) $p$ = 0.620(8) measured at $T$ = 1.4 K. The neutron scattering measurements were performed with incident neutron wavelength of 6 angstrom. Intensities were normalized to an absolute unit by comparing them to the (0,0,2) nuclear Bragg peak intensity.

**Fig. 3** (a) $Q$-dependence of the elastic magnetic scattering intensity measured for various values of $p$ at 1.4 K, except for $p = 0.459(5)$ at $T = 0.27$ K. Nonmagnetic background was determined at 20 K and subtracted. Solid lines are fits to a simple Lorentzian. Dashed lines and arrows represent the fitted Full-Width-of-Half-Maximum (FWHM) and peak positions, respectively. (b) The peak position and the power static spin correlation length that were obtained from the fits, are plotted as a function of $p$. (c) The fraction of the contribution from the single Lorentzian to the dynamic susceptibility, as shown in Fig. 4, is plotted as a function of $p$.

**Fig. 4** The imaginary part of the dynamic susceptibility at low energies was obtained from the inelastic neutron scattering intensities (see Supporting Online Material) at several different temperatures for various values of $p$: (a) $p = 0.968(6)$, (b) $p = 0.917(9)$, (c) $p = 0.844(8)$, (d) $p = 0.777(6)$, (e) $p = 0.620(8)$ and (f) $p = 0.459(5)$. The solid, dotted and dashed lines in (a)-(f) are explained in the text.

Fig. 1.

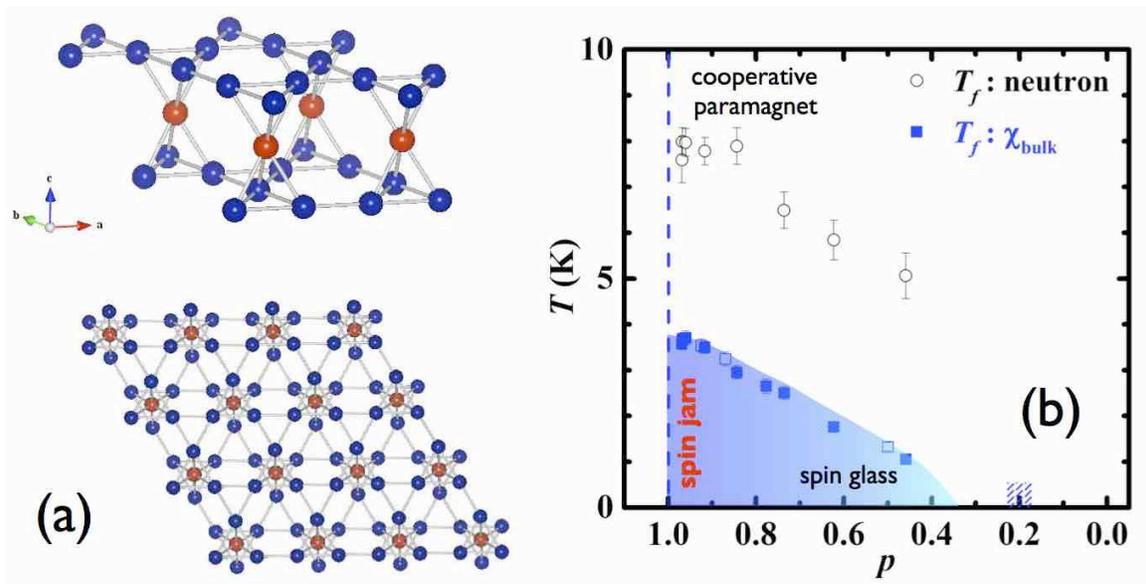

Fig. 2

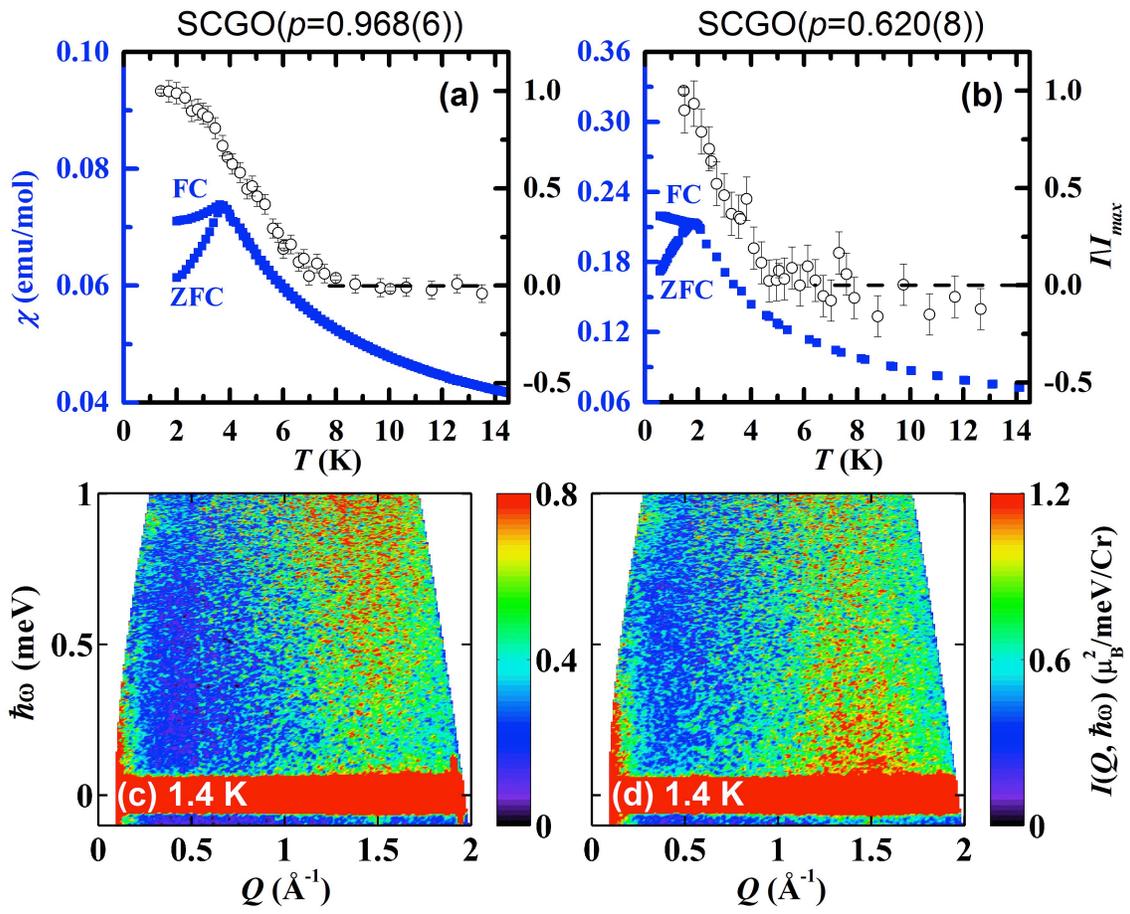

Fig. 3

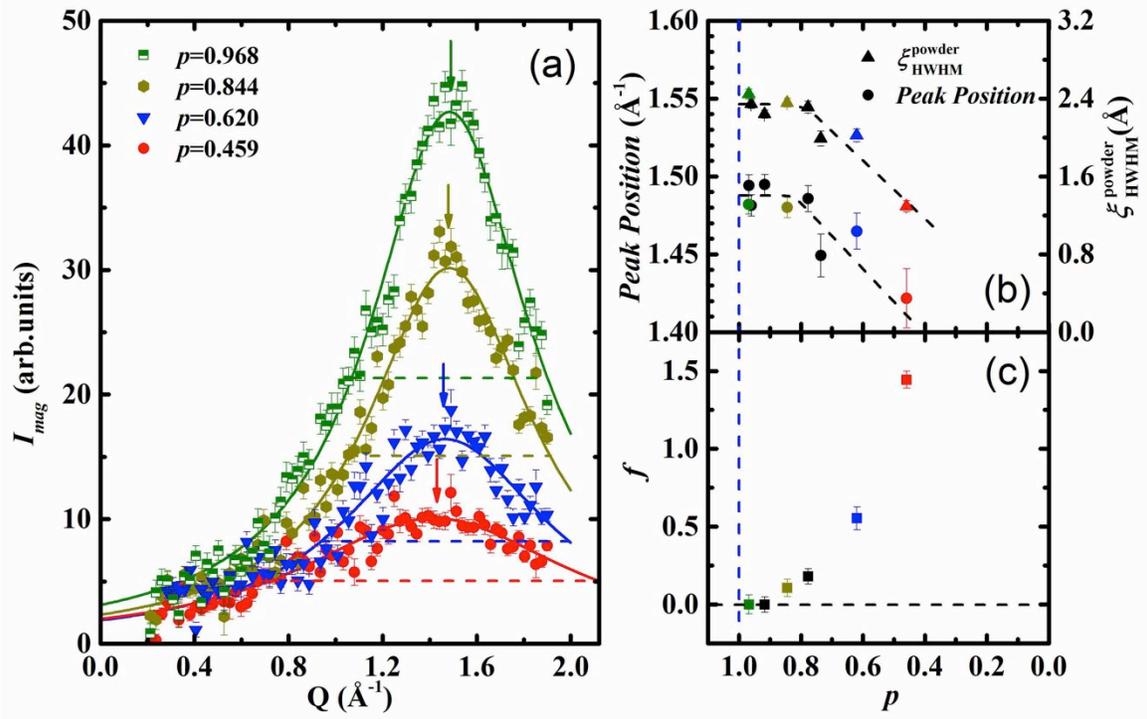

Fig. 4

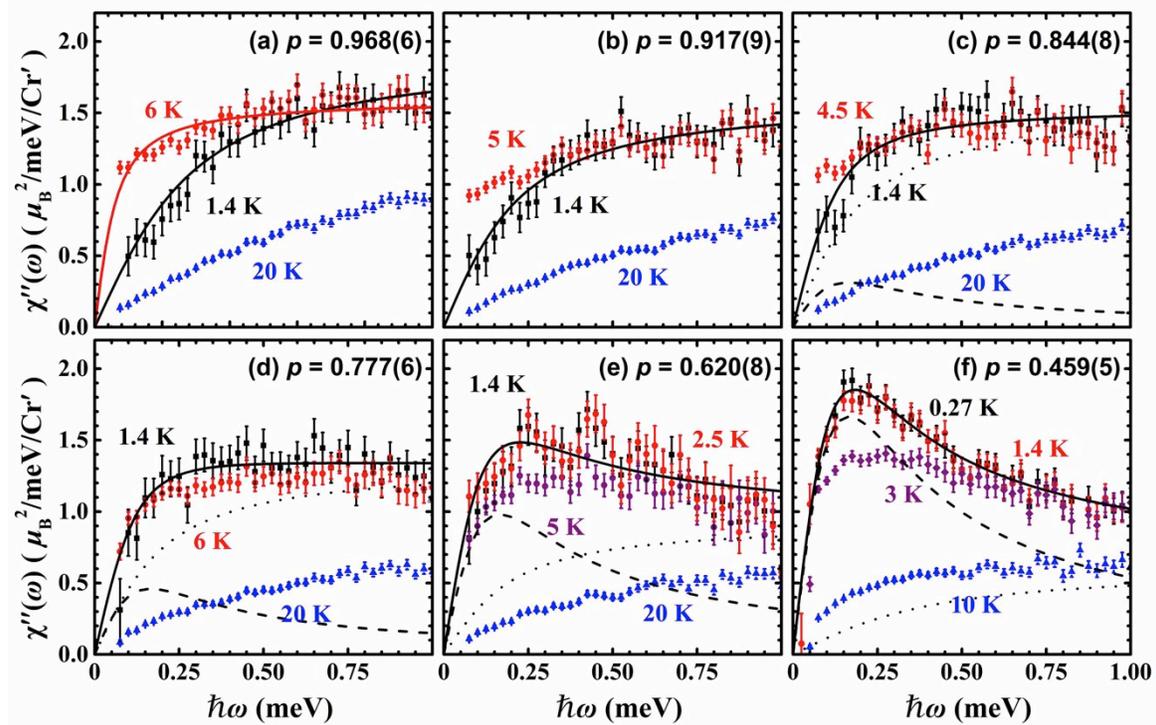

# Supporting Online Material

**Methods and materials**

Polycrystalline samples of SCGO($p$) with twelve different values of $p$ = 0.968(6), 0.960(9), 0.925(nominal), 0.917(9), 0.87(nominal), 0.844(8), 0.777(6), 0.736(6), 0.620(8), 0.5(nominal), 0.459(5) and 0.228(5) were made using standard solid-state synthesis techniques at the University of Virginia. Stoichiometric quantities of $SrCO_3$, $Ga_2O_3$ and $Cr_2O_3$ were mixed, ground and pelletized. Then, the pellets were heated in air in dense $Al_2O_3$ crucibles at 1400 °C for 96 h, with three intermediate grindings. The samples were then annealed at 1000 °C for 7 days to promote cation ordering. DC magnetic susceptibilities were measured using the Quantum design MPMS-XL7 with a Helium-3 insert in the temperature range from 0.5 K to 14 K at ETH Zurich, while similar measurements were done from 2 K to 300 K at Kyoto University. Crystal structure and the actual Cr concentrations were determined by neutron diffraction. Neutron powder diffraction measurements were performed using BT1 diffractometer at the NIST Center for Neutron Research (NCNR) located in Gaithersburg, Maryland, with the neutron wavelength of $\lambda = 1.54$ Å at room temperature for SCGO($p$) samples with $p$ = 0.968(6), 0.960(9), 0.917(9), 0.844(8), 0.777(6), 0.736(6), 0.620(8), 0.459(5) and 0.228(5).

Time-of-flight neutron scattering measurements were performed using the Disk-Chopper-Spectrometer (DCS) at NCNR with $\lambda = 6$ Å. SCGO($p$) samples with $p$ = 0.968(6), 0.960(9), 0.917(9), 0.844(8), 0.777(6), 0.736(6) and 0.620(8) were measured in a standard DCS ILL orange cryostat and $p$ = 0.459(5) was measured with a Helium-3 insert. The low energy inelastic magnetic neutron scattering spectrum was measured at various temperatures, $T$, around $T_f$ for the samples. For each SCGO($p$) sample, an energy dependent background was determined by imposing the detailed balance constraint

$S(-\hbar\omega, T) = e^{-\frac{\hbar\omega}{k_B T}} S(\hbar\omega, T)$ where $k_B$ is the Boltzmann constant, and it was subtracted from the data to obtain the magnetic contribution. Then, the imaginary part of the dynamic susceptibility was derived from the fluctuation dissipation theorem, $\chi''(\hbar\omega) = \pi \left\{ 1 - e^{-\frac{\hbar\omega}{k_B T}} \right\} S(\hbar\omega)$.

**Bulk magnetic susceptibility and Curie-Weiss fitting.**

Fig. S1 shows the reciprocal susceptibility as a function of temperature ($1/\chi$ vs $T$) measured in a 0.1 Tesla field over a temperature range of 2 – 300 K. The Curie-Weiss temperatures ($\Theta_{CW}$) were extracted by fitting the $1/\chi$-$T$ curves at $T > 150$ K with Curie-Weiss law ($\chi=C/(T-\Theta_{CW})$). The $\Theta_{CW}$ values range from -72 to -504 K, reflecting the strong anti-ferromagnetic interactions in the system. As shown in the Fig. S1 inset, $|\Theta_{CW}|$ decreases gradually with decreasing Cr concentration from $p = 0.960(9)$ to $0.228(5)$.

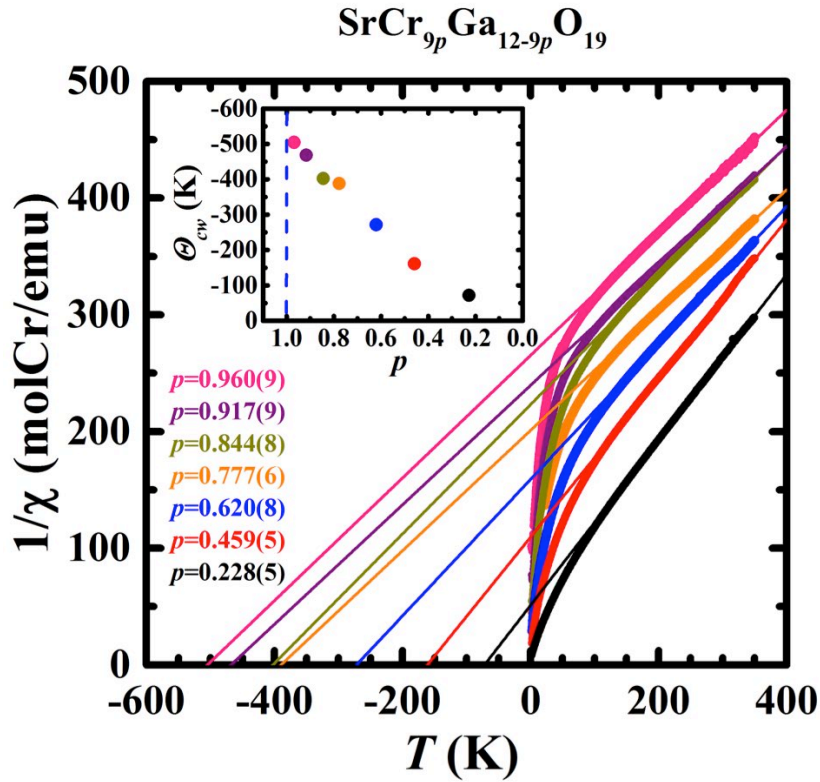

**Figure S1.** Reciprocal bulk magnetic susceptibility of SCGO ($0.228(5) \leq p \leq 0.960(9)$). The straight lines are the linear fitting at $T > 150$ K. The inset shows the $\Theta_{CW}$ values obtained by Curie-Weiss fitting.

**Neutron powder diffraction refinement.**

Figure S2 shows the neutron powder diffraction data of SCGO ($p$ = 0.968(6), 0.844(8), 0.620(8) and 0.495(5)). The lines are the Rietveld refinement results of nuclear structure performed using General Structure Analysis System (GSAS) (*S1*). The optimal parameters for their crystal structures are listed in Tables S1 to S4, which are consistent with the previously published structure (*S2,S3*).

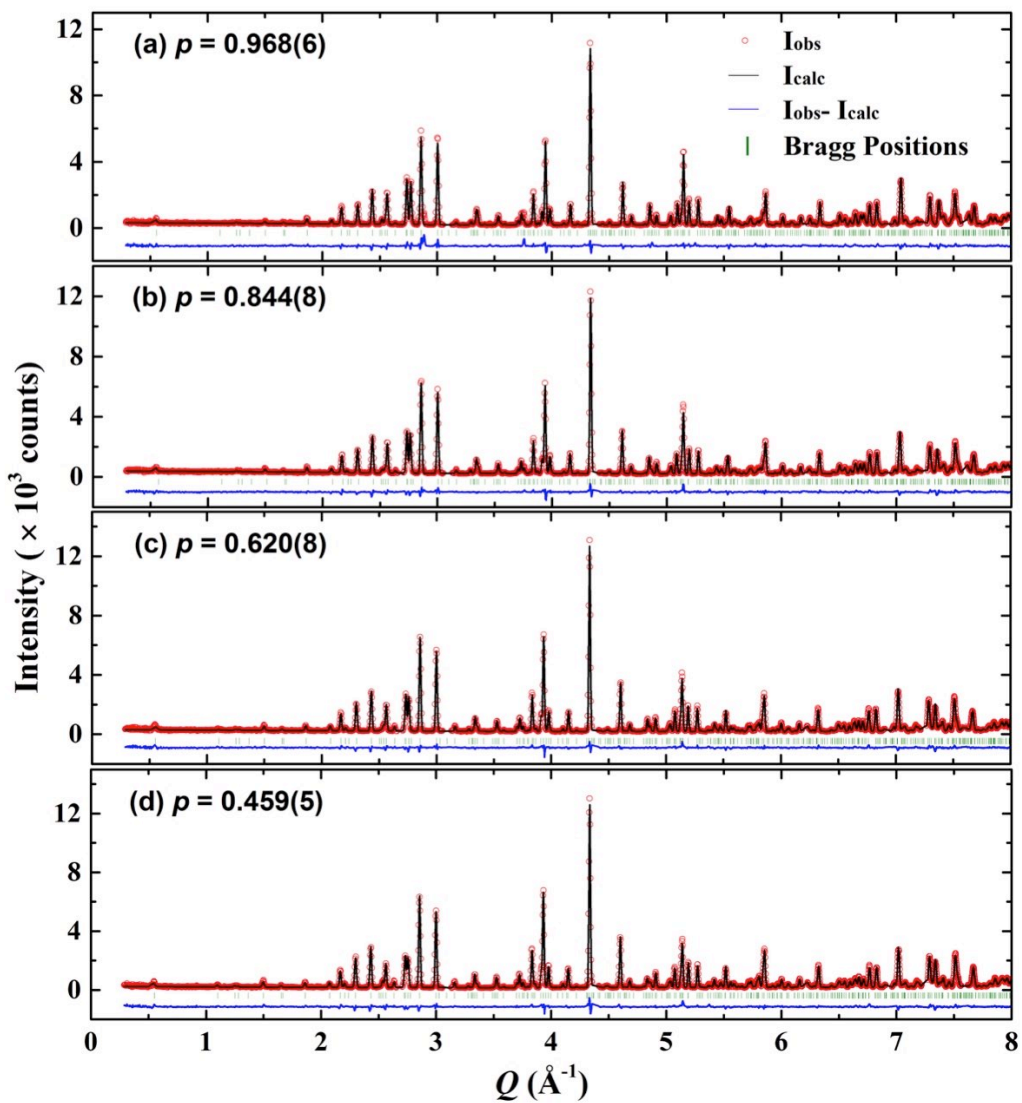

**Figure S2.** Neutron powder diffraction data (open circles) and Rietveld refinement patterns

(solid curves) for SCGO(p) samples: (a) $p = 0.968(6)$, (b) $p = 0.844(8)$, (c) $p = 0.620(8)$ and (d) $p = 0.459(5)$. The green vertical marks represent the position of Bragg peaks, and the blue solid line at the bottom corresponds to the difference between the observed and calculated intensities. The $p$ values shown in each panel were obtained from the Rietveld refinement.

**Table S1.** $p = 0.968(6)$. Positions within space group $P6_3/mmc$ and occupancies per formula unit (f.u.) of atoms in SCGO(p) at $T = 298$ K as determined by Rietveld analysis of the data shown in Fig. S2 using the GSAS. The lattice parameters are $a = 5.7948(3)$ Å and $c = 22.6625(2)$ Å. Isotropic Debye-Waller factors, $\exp(-\langle u^2 \rangle Q^2)$, were used where $\langle u^2 \rangle$ is the mean squared displacement. The resulting overall reduced $\chi^2 = 2.578$

| Site | $x$ | $y$ | $z$ | $n$/f.u. | $\sqrt{\langle u^2 \rangle}$/Å |
|---|---|---|---|---|---|
| Sr 2$d$ | 2/3 | 1/3 | 1/4 | 1 | 0.0086(8) |
| Ga 2$a$ | 0 | 0 | 0 | 0.036(18) | 0.0041(4) |
| Cr 2$a$ | 0 | 0 | 0 | 0.964(18) | 0.0041(4) |
| Ga 2$b$ | 0 | 0 | 1/4 | 1 | 0.0128(7) |
| Ga 4$f_{iv}$ | 1/3 | 2/3 | 0.0275(1) | 2 | 0.0062(4) |
| Ga 4$f_{vi}$ | 1/3 | 2/3 | 0.1911(2) | 0.056(30) | 0.0041(4) |
| Cr 4$f_{vi}$ | 1/3 | 2/3 | 0.1911(2) | 1.944(30) | 0.0041(4) |
| Ga 12$k$ | 0.1685(3) | 0.3370(6) | −0.1081(1) | 0.192(60) | 0.0041(4) |
| Cr 12$k$ | 0.1685(3) | 0.3370(6) | −0.1081(1) | 5.808(60) | 0.0041(4) |
| O 4$e$ | 0 | 0 | 0.1520(2) | 2 | 0.0080(7) |
| O 4$f$ | 1/3 | 2/3 | −0.0546(2) | 2 | 0.0019(5) |
| O 6$h$ | 0.1813(3) | 0.3635(6) | 1/4 | 3 | 0.0086(4) |
| O 12$k$ | 0.1556(2) | 0.3112(4) | 0.0529(1) | 6 | 0.0037(3) |
| O 12$k$ | 0.5055(2) | 0.0111(4) | 0.1507(1) | 6 | 0.0059(3) |

**Table S2.** $p = 0.844(8)$. The lattice parameters at 298 K are $a = 5.79447(3)$ Å and $c = 22.7120(2)$ Å. The resulting overall reduced $\chi^2 = 2.290$.

| Site | $x$ | $y$ | $z$ | $n$/f.u. | $\sqrt{\langle u^2 \rangle}$/Å |
|---|---|---|---|---|---|
| Sr 2$d$ | 2/3 | 1/3 | 1/4 | 1 | 0.0098(7) |
| Ga 2$a$ | 0 | 0 | 0 | 0.131(17) | 0.0045(3) |
| Cr 2$a$ | 0 | 0 | 0 | 0.869(17) | 0.0045(3) |

| Site | x | y | z | n/f.u. | $\sqrt{\langle u^2 \rangle}$/Å |
|---|---|---|---|---|---|
| Ga 2b | 0 | 0 | 1/4 | 1 | 0.0141(7) |
| Ga 4$f_{iv}$ | 1/3 | 2/3 | 0.0272(1) | 2 | 0.0054(4) |
| Ga 4$f_{vi}$ | 1/3 | 2/3 | 0.1906(2) | 0.321(28) | 0.0044(4) |
| Cr 4$f_{vi}$ | 1/3 | 2/3 | 0.1906(2) | 1.679(28) | 0.0044(4) |
| Ga 12k | 0.1686(3) | 0.3374(5) | −0.1089(1) | 0.948(54) | 0.0045(4) |
| Cr 12k | 0.1686(3) | 0.3374(7) | −0.1089(1) | 5.052(54) | 0.0045(4) |
| O 4e | 0 | 0 | 0.1514(1) | 2 | 0.0096(6) |
| O 4f | 1/3 | 2/3 | −0.0551(1) | 2 | 0.0019(5) |
| O 6h | 0.1809(2) | 0.3618(5) | 1/4 | 3 | 0.0094(5) |
| O 12k | 0.1555(2) | 0.3109(3) | 0.0525(1) | 6 | 0.0043(3) |
| O 12k | 0.5054(2) | 0.0108(4) | 0.1504(1) | 6 | 0.0049(3) |

**Table S3.** $p$ = 0.620(8). The lattice parameters at 298 K are $a$ = 5.79676(3) Å and $c$ = 22.7541(2) Å. The resulting overall reduced $\chi^2 = 2.195$.

| Site | x | y | z | n/f.u. | $\sqrt{\langle u^2 \rangle}$/Å |
|---|---|---|---|---|---|
| Sr 2d | 2/3 | 1/3 | 1/4 | 1 | 0.0107(7) |
| Ga 2a | 0 | 0 | 0 | 0.298(17) | 0.0047(3) |
| Cr 2a | 0 | 0 | 0 | 0.702(17) | 0.0047(3) |
| Ga 2b | 0 | 0 | 1/4 | 1 | 0.0159(7) |
| Ga 4$f_{iv}$ | 1/3 | 2/3 | 0.0276(1) | 2 | 0.0044(4) |
| Ga 4$f_{vi}$ | 1/3 | 2/3 | 0.1905(1) | 0.790(28) | 0.0047(3) |
| Cr 4$f_{vi}$ | 1/3 | 2/3 | 0.1905(1) | 1.210(28) | 0.0047(3) |
| Ga 12k | 0.1682(2) | 0.3366(4) | −0.1093(1) | 2.328(66) | 0.0047(3) |
| Cr 12k | 0.1682(2) | 0.3366(4) | −0.1093(1) | 3.672(66) | 0.0047(3) |
| O 4e | 0 | 0 | 0.1504(1) | 2 | 0.0102(6) |
| O 4f | 1/3 | 2/3 | −0.0547(1) | 2 | 0.0019(5) |
| O 6h | 0.1811(3) | 0.3622(5) | 1/4 | 3 | 0.0089(5) |
| O 12k | 0.1556(2) | 0.3112(3) | 0.0526(1) | 6 | 0.0046(3) |
| O 12k | 0.5047(2) | 0.0094(4) | 0.1502(1) | 6 | 0.0051(3) |

**Table S4.** $p = 0.459(5)$. The lattice parameters at 298 K are $a = 5.79566(3)$ Å and $c = 22.7778(2)$ Å. The resulting overall reduced $\chi^2 = 2.448$.

| Site | $x$ | $y$ | $z$ | $n$/f.u. | $\sqrt{\langle u^2 \rangle}$/Å |
|---|---|---|---|---|---|
| Sr 2$d$ | 2/3 | 1/3 | 1/4 | 1 | 0.0108(8) |
| Ga 2$a$ | 0 | 0 | 0 | 0.431(16) | 0.0036(3) |
| Cr 2$a$ | 0 | 0 | 0 | 0.569(16) | 0.0036(3) |
| Ga 2$b$ | 0 | 0 | 1/4 | 1 | 0.0159(7) |
| Ga 4$f_{iv}$ | 1/3 | 2/3 | 0.0276(1) | 2 | 0.0050(4) |
| Ga 4$f_{vi}$ | 1/3 | 2/3 | 0.1897(1) | 1.134(13) | 0.0037(3) |
| Cr 4$f_{vi}$ | 1/3 | 2/3 | 0.1897(1) | 0.866(13) | 0.0037(3) |
| Ga 12$k$ | 0.1683(2) | 0.3366(4) | −0.1094(1) | 3.306(30) | 0.0037(3) |
| Cr 12$k$ | 0.1683(2) | 0.3366(4) | −0.1094(1) | 2.694(30) | 0.0037(3) |
| O 4$e$ | 0 | 0 | 0.1501(1) | 2 | 0.0097(6) |
| O 4$f$ | 1/3 | 2/3 | −0.0545(1) | 2 | 0.0024(5) |
| O 6$h$ | 0.1812(3) | 0.3623(5) | 1/4 | 3 | 0.0093(5) |
| O 12$k$ | 0.1554(2) | 0.3106(3) | 0.0526(7) | 6 | 0.0054(3) |
| O 12$k$ | 0.5047(2) | 0.0094(4) | 0.1500(1) | 6 | 0.0061(3) |

**Low temperature magnetic susceptibility and elastic neutron intensity.**

The low temperature magnetic susceptibilities were measured in a field of 0.01 Tesla in the temperature range from 0.5 K to 14 K, as shown in Fig. S3 and Fig. S4 (blue squares). The susceptibility increases with decreasing temperature, exhibiting a low-temperature upturn. With further decreasing temperature, a divergence between the filed cooled (FC) and zero-field cooled (ZFC) susceptibility, which is the typical behavior of spin-glass transition, was observed for all the six samples. The freezing temperature $T_f$ was determined as the temperature of the maximum susceptibility for the ZFC curve, where the divergence between FC and ZFC susceptibility also occurs. $T_f$ decreases with decreasing Cr concentration. Above $T_f$, no difference was observed between FC and ZFC susceptibilities for all the samples.

The wave vector averaged magnetic elastic neutron scattering intensities normalized to the maximum intensity as a function of temperature are also shown in Fig. S3 (open circles). Neutron scattering data were obtained using the cold-neutron DCS at the NCNR with $\lambda = 6.0$ Å. The temperature of the onset of magnetic neutron intensity is higher than the freezing temperature observed in bulk magnetic susceptibility data, and decreases with decreasing Cr concentrations exhibiting a similar behavior to the one observed in the susceptibility data.

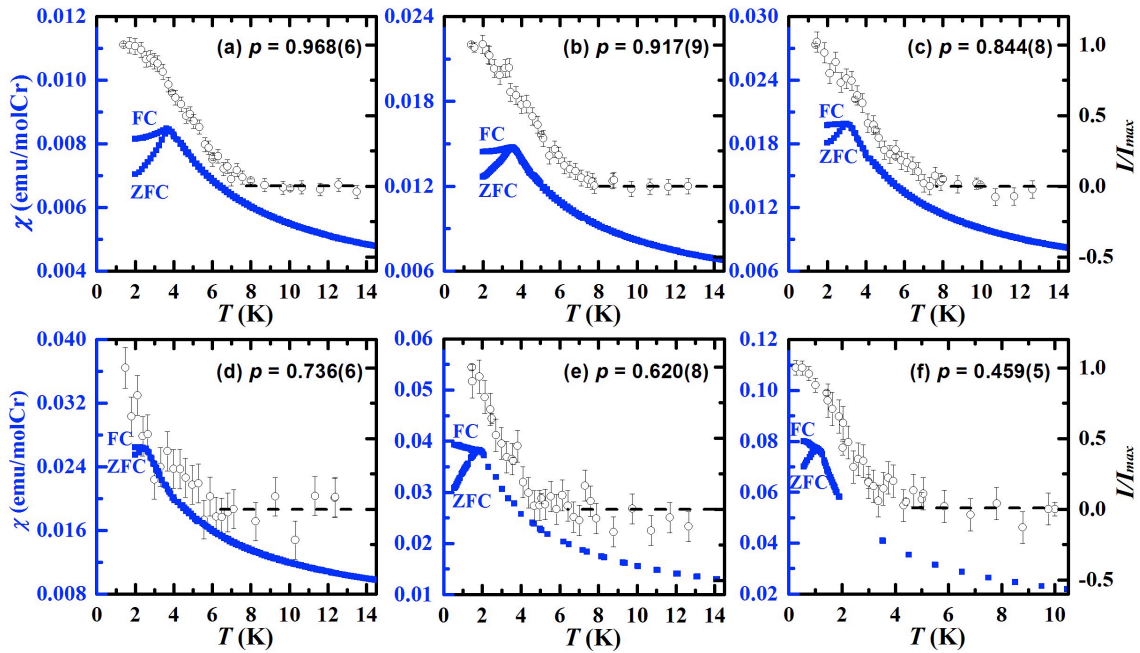

**Figure S3.** Temperature-dependent magnetic susceptibility (blue solid squares) in $H = 0.01$ T and the wave vector averaged magnetic elastic neutron scattering intensity (black open circles) for SCGO ($0.459(5) \leq p \leq 0.968(6)$) samples: (a) $p = 0.968(6)$, (b) $p = 0.917(9)$, (c) $p = 0.844(8)$, (d) $p = 0.736(6)$, (e) $p = 0.620(8)$ and (f) $p = 0.459(5)$. The black dash lines indicate the background.

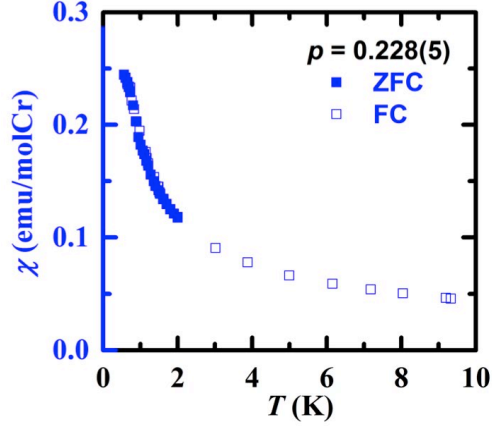

**Figure S4.** Bulk magnetic susceptibility (blue solid squares) as a function of temperature, measured for SCGO ($p = 0.228(5)$) under a weak external magnetic field of $H = 0.01$ T.

**Contour maps of inelastic neutron scattering data.**

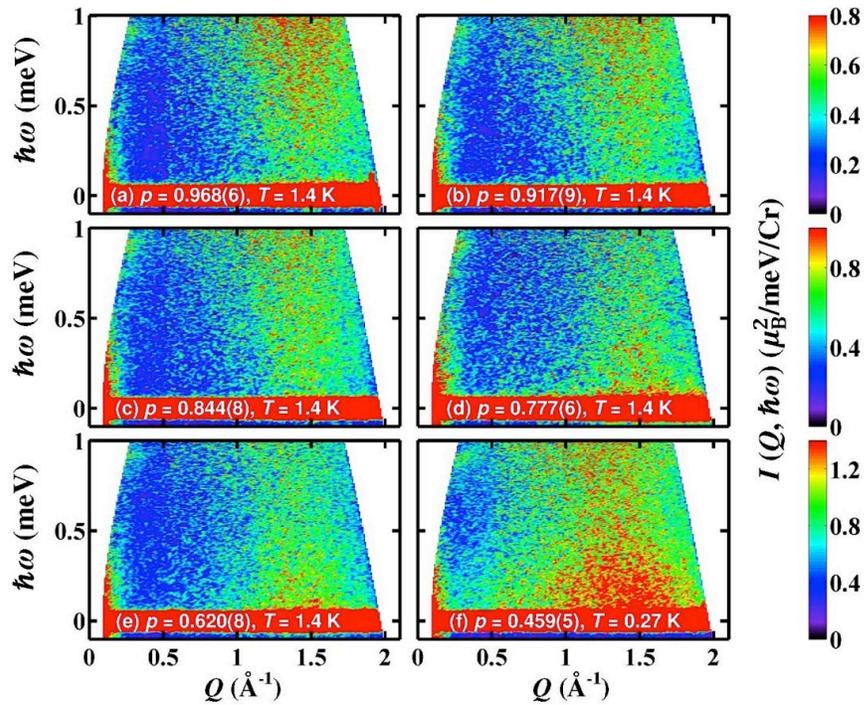

**Figure S5.** Contour maps of neutron scattering intensity $I(Q,\omega)$ as a function of energy transfer $\hbar\omega$ and momentum transfer $Q$ for (a) $p = 0.968(6)$ at $T = 1.4$ K, (b) $p = 0.917(9)$ at $T = 1.4$ K, (c) $p = 0.844(8)$ at $T = 1.4$ K, (d) $p = 0.777(6)$ at $T = 1.4$ K, (e) $p = 0.620(8)$ at $T = 1.4$ K and (f) $p = 0.459(5)$ at $T = 0.27$ K with $\lambda = 6$ Å. Intensities were normalized to an absolute unit by scale factors obtained from the (0,0,2) nuclear Bragg peak intensity.

Figure S5 shows the low energy inelastic magnetic neutron scattering spectrum measured in the frozen states of several different SCGO( p) samples. In the frozen state of $p = 0.968(6)$, the inelastic neutron scattering intensity $I(Q, \hbar\omega)$ is weak at very low energies below 0.25 meV and gets stronger as $\hbar\omega$ increases. In contrast, in the frozen state of $p = 0.459(5)$, $I(Q, \hbar\omega)$ is strong at very low energies below 0.25 meV and gets weaker as $\hbar\omega$ increases. This stark difference hints that the frozen states of $p = 0.968(6)$ and $0.459(5)$ are different in nature. The change of $I(Q, \hbar\omega)$ at very low energies below 0.25 meV is gradual with decreasing $p$ and manifest when $p = 0.777(6)$, indicating that a transition to a different frozen state occurs between $p = 0.776(6)$ and $0.844(8)$.

**S6, Spin wave dispersion.**

The spin wave calculations were done for the SCGO at the clean limit, $p=1$. This was done by expanding around classically ordered ground state configuration to harmonic order, using the standard Holstein-Primakoff method (*S4*).

The chosen states are coplanar spin configurations that are described in the inset in Fig. S6. The spin wave dispersion is plotted along the (H,H,1.8) momentum direction for various values of $\alpha = \cos^{-1}\left(\frac{J'/J+1}{2}\right)$ (*S5*), going from $\alpha = 0$ for the ideal SCGO to $\alpha = \pi/3$ for isolated kagome. The spin wave analysis shows a flat band of 9 zero modes, indicating a large classical degeneracy.

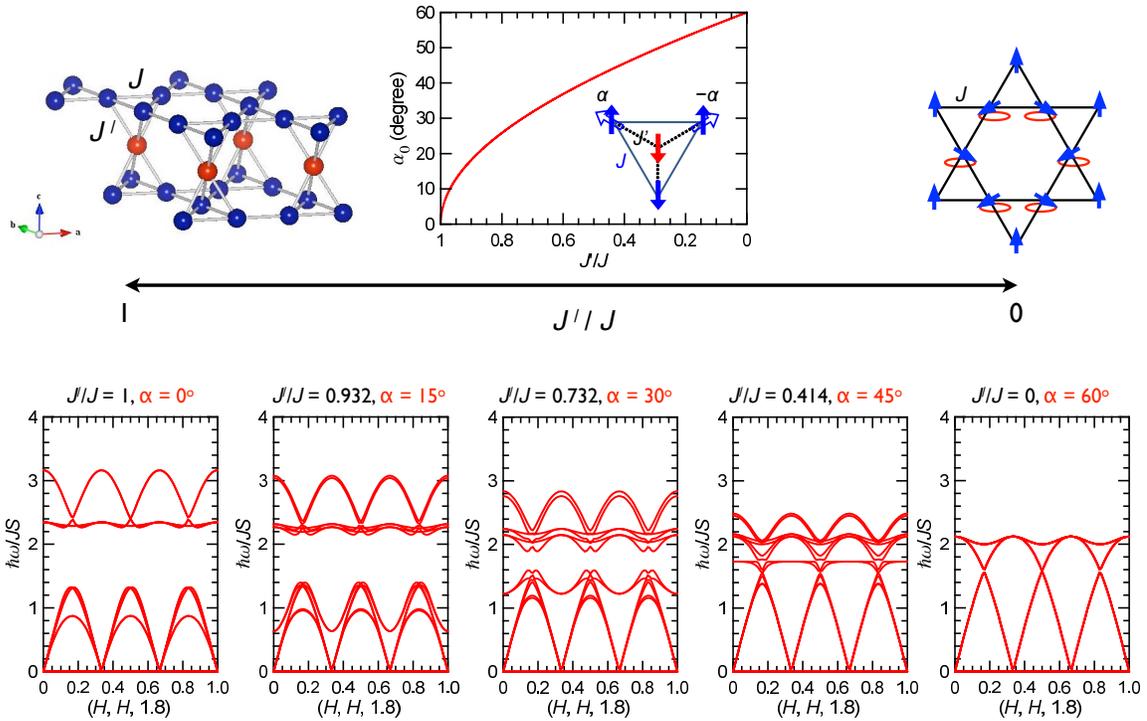

**Figure S6.** Spin wave dispersion. The ideal SCGO lattice and the kagome may be viewed as limits as $J'/J$ goes from 1 to 0. Ordered classical coplanar ground state configurations may be obtained from the ideal case by collectively rotating a subset of the spins as is shown in the inset and Ref. (*S5*).